\documentclass[preprint,12pt]{elsarticle}
\usepackage{epsf,epsfig}
\usepackage[psamsfonts]{amssymb}
\usepackage{amsmath}
\usepackage{bm}
\usepackage{graphicx}
\usepackage{graphicx,epstopdf}
\usepackage{caption}
\usepackage{subcaption}
\usepackage{graphicx}

\newcommand{\be}{\begin{equation}}
\newcommand{\ee}{\end{equation}}
\newcommand{\bea}{\begin{eqnarray}}
\newcommand{\eea}{\end{eqnarray}}
\newcommand{\h}{\hspace{0.30 cm}}

\newcommand{\n}{\nonumber} %---------------------------------------------------------------------------------- \begin{document}

\begin{document}

\begin{frontmatter}
\title{Free particle and isotropic harmonic oscillator on a spheroidal surface: the Higgs-like approach}
\author{Ali Mahdifar $^{1,2}$}

\ead{mahdifar$_{-}$a@sci.sku.ac.ir}

\author{Ehsan Amooghorban $^{1,3}$}

\address{ $^1$ Department of Physics, Faculty of Basic Sciences, Shahrekord University, Shahrekord 88186-34141, Iran.}
\address{$^2$ Department of Physics, Faculty of Science, University of Isfahan, Hezar Jerib, Isfahan, 81746-73441,Iran. }
\address{$^3$ Nanotechnology Research Center, Shahrekord University, Shahrekord 88186-34141, Iran.}

\begin{abstract}
In this paper, we investigate the dynamics of both free particle and isotropic harmonic oscillator constrained to move on a spheroidal surface using two consecutive projections: a projection onto a sphere surface followed by the gnomonic projection onto a tangent plane to the spheroid.
We obtain the Hamiltonian of the aforementioned systems in terms of the Cartesian coordinates
of the tangent plane and then quantize it in the standard way.
It is shown that the effect of non-sphericity of the surface can be treated as the appearance of an effective potential.
By using the perturbation theory up to the first order in second eccentricity of the spheroid, we approximately calculate the eigenfunctions and eigenvalues of the free particle, as well as the isotropic harmonic oscillator on the spheroidal surface.
We find that the deviation from the sphericity plays an important role in splitting the energy levels of the isotropic oscillator on a sphere, and lifting the
degeneracy.
\end{abstract}


\end{frontmatter}

\section{Introduction}

The study of quantum mechanical systems in the curved spaces is an interesting and important problem due to relation with Einstein's theory of gravitation in general relativity \cite{Birrell1994,Parker2009}, and its potential application in many different fields of physics from the
the quantum Hall effect \cite{Bracken2007} and the quantum dots \cite{Gritsev2001,Bulaev2004} to the coherent state quantization \cite{Gazeau2004,Mahdifar2006}.
At a practical level, due to technological developments in nanotechnology, photonics and plasmonics, it made possible to fabricate complex heterostructures with new properties as a result of confinement of electrons or light on curved surfaces.
In addition, one can study the propagation of electromagnetic waves constrained to a film waveguide and a curved optical fiber in terms of the models which are founded on quantum mechanics on curved spaces. This comes from the analogy between the Helmholtz's and the time-independent Schr\"{o}dinger equation.
One can also treat these systems as analog models of general relativity~\cite{Leonhardt2006,Tavakoli2009}.
Moreover, these studies are also of theoretical interest, because it is not clear how to generalize the well defined quantum processes such as the quantization from the flat space to the curved spaces, even though the curvature is constant but different from zero.

The first attempt to study quantum mechanical systems in the spaces of constant curvature was given by Schr\"{o}dinger, who in the framework of the factorization method investigated the hydrogen atom in a three-dimensional sphere~\cite{Schr}. Almost at the same time, Infeld and Shild studied the Schr\"{o}dinger equation for the same problem in an open hyperbolic universe~\cite{Infeld1945}.
To the best of our knowledge, there are three known approaches to treat non-relativistic quantum mechanics in the spaces of constant curvature:
The first one is the Noether quantization, in which the Killing vector fields
provide the Noether momenta for the system. The Hamiltonian is  written in terms of these momenta and subsequently the quantization process is applied to the components of the Noether momenta~\cite{deh 9, deh 10,Bracken2014}.
Another method is the so-called thin layer quantization, which was introduced in
the seminal papers~\cite{Jensen1971,deh 1,da Costa1982,deh}. In this approach, the 2D surface is embedded into the larger 3D Euclidean space and then the dimensional
reduction in the Schr\"{o}dinger equation is achieved by introducing an effective potential.
The third method is the Higgs approach \cite{Higgs1979,Leemon1979}. In this approach, dynamical symmetries are worked out in a spherical
geometry. In this manner, the Hamiltonian of the system is associated with the Casimir operators of the Lie algebra of these dynamical symmetries, and then the energy eigenvalues are obtained by the eigenvalues of the Casimir operators.

Higgs~\cite{Higgs1979} and Leemon~\cite{Leemon1979} investigated the non-relativistic motion of a particle on a $N$-dimensional sphere (embedded in the Euclidean $(N + 1)$-dimensional space) under the action of the force field with specific centrally symmetric potential. In the cases of flat Euclidean space where the curvature vanishes, these central potentials reduce to the usual isotropic oscillator and Coulomb potentials. Higgs described the motion on a sphere by means of the so-called gnomonic projection which is a projection onto the tangent plane from the center of the sphere in the embedding space.

The advantage of this projection is to transform the uniform motion of free particle on a great circle into a rectilinear, but non-uniform motion on the tangent plane. This means that the projected free particle orbits are the same as those in the flat Euclidean space, whereas the effect of curvature is included in the velocity of the projected motion. Amazingly, this feature survives even in the presence of a central force derived from a potential $V(r)$, {\rm i.e.}, the dynamical symmetries in a sphere geometry are the same as those in tangent space.

One of authors of this paper has recently analyzed the dynamical behavior of a two-dimensional isotropic harmonic oscillator constrained to a spherical surface with a time-dependent radius~\cite{Mirza}. It is a remarkable fact that time variations in the sphere radius resulted in
a minimal-coupling interaction Hamiltonian. Within a simple golden rule calculation, it has been shown that the isotropic harmonic oscillator on the sphere is excited through appropriate frequencies of the background fluctuations.

More recently, we have investigated the classical and the quantum mechanical treatment
of a damped particle on a sphere under the action of a conservative central force~\cite{DisHig}.
In this approach, the dissipation and the fluctuation effects are introduced to our formalism by interacting the main system with a reservoir which modeled by a continuum of three dimensional harmonic oscillators.
We found that the dynamics of the dissipative Higgs
model is determined by an effective geometry-induced susceptibility which included the extrinsic geometry of
the physical space, and a noise operator arises in the result of absorption.
Due to use the gnomonic projection, the anisotropy appears in the projected susceptibility.
In particular, it has been shown that appreciable probabilities for transition are
possible only if the transition and reservoir's oscillators frequencies
to be nearly on resonance.

In the present contribution, our main purpose is to generalize the Higgs model from the spherical to a spheroidal surface. A two-dimensional spheroid is a quadric surface obtained by rotating an ellipse about one of its principal axes (an ellipsoid with two equal semi-diameters).
For this purpose, we investigate the motion of a non-relativistic particle on a $2$-dimensional spheroid (embedded in $\mathbb{R}^3$) under the isotropic harmonic oscillator potential. This central potential reduces to the isotropic oscillator of a Euclidean geometry when the curvature of the spheroid goes to zero.

The paper is organized as follows.
In section \ref{QHO}, by making use a projection from the spheroidal to the spherical space and then the gnomonic projection onto the tangent plane, we obtain the Hamiltonian of an isotropic harmonic oscillator confined to a spheroidal background in terms of the Cartesian coordinates of the tangent plane to the spheroid.  We subsequently quantize the aforementioned Hamiltonian by replacing classical position and momentum by related operators.
In section~\ref{FPS}, first we calculate the eigenfunctions and eigenvalues of a free particle on a sphere and then, by using the perturbation theory up to the first order in second eccentricity of the spheroid, we approximately determine the eigenvalues and the eigenfunctions of a free particle on the spheroidal surface.
In section \ref{IHO}, the eigenvalues and the eigenfunctions for an isotropic harmonic oscillator problem on the spheroidal surface are approximately derived.
Finally, the summary and concluding remarks are given in section \ref{summary}.

\section{Quantum isotropic harmonic oscillator on a spheroid}\label{QHO}

\begin{figure}[t]
\includegraphics[width=0.5\columnwidth]{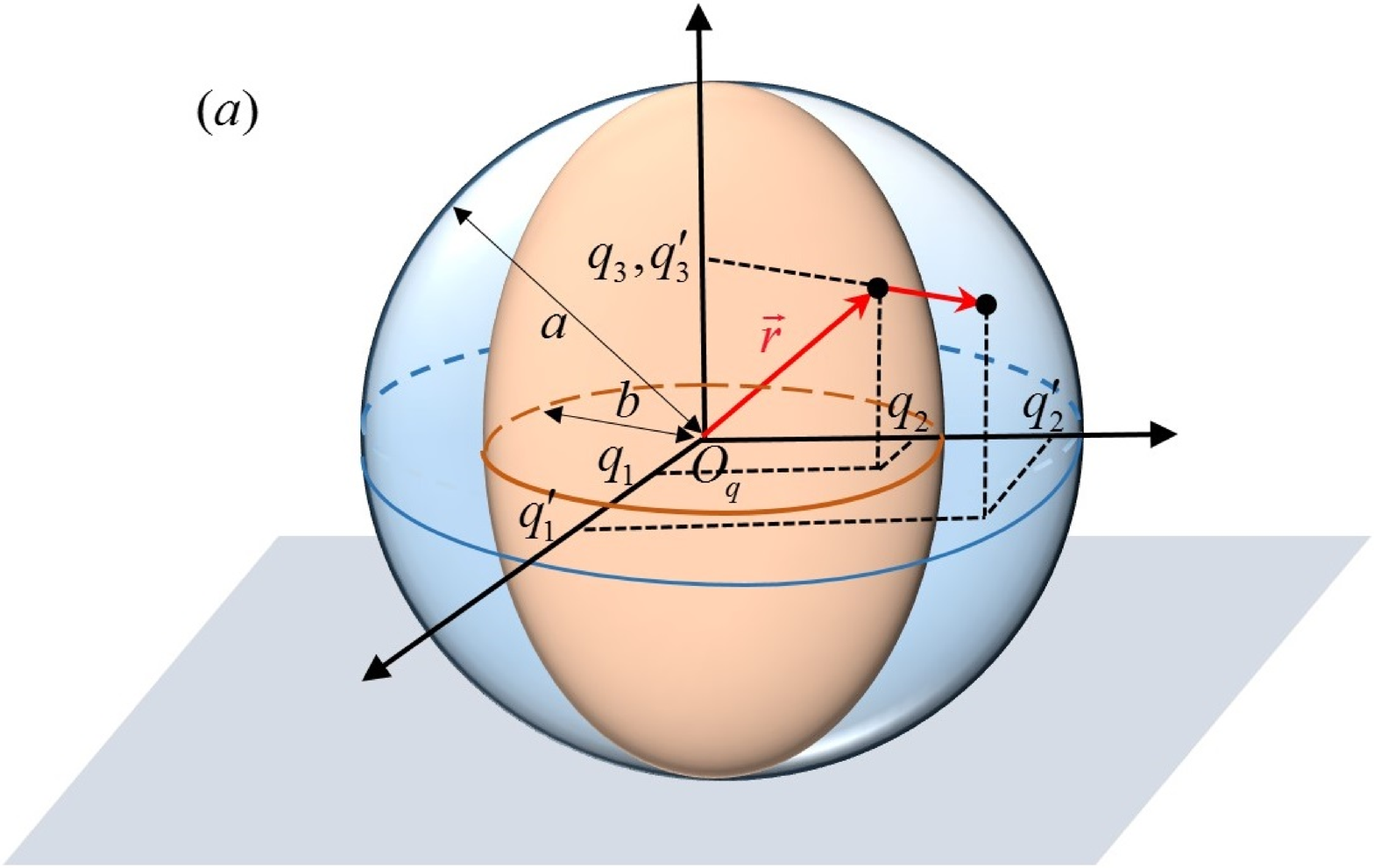}
\includegraphics[width=0.5\columnwidth]{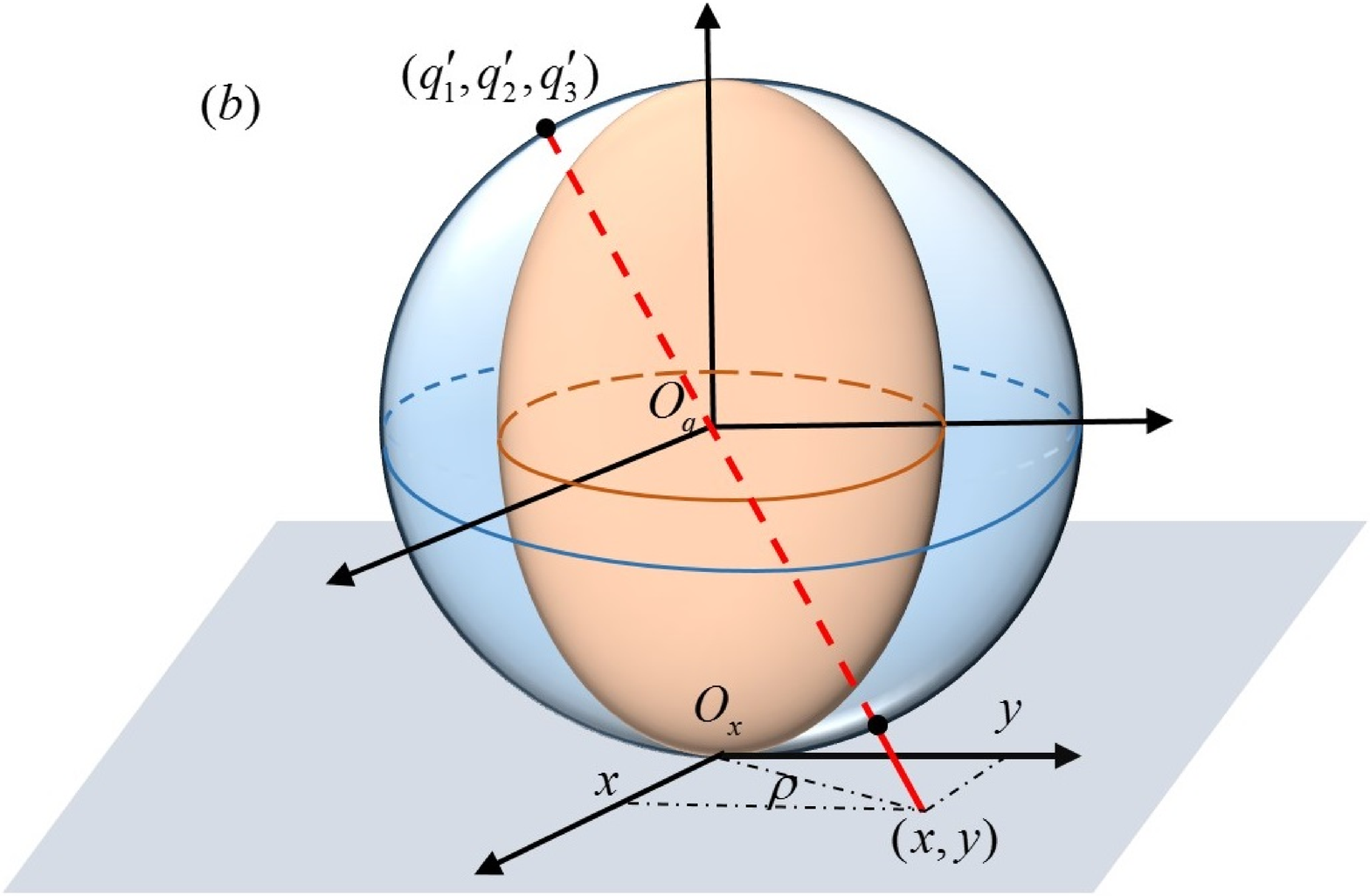}
\caption{Coordinate systems and projection from a spheroid with polar radius $a$ and equatorial radius $b$ onto a plane. (a) A given point on the spheroid with coordinates $(q_1,q_2,q_3)$ is projected onto
the surface of an enclosed-sphere having radius $a$ with coordinates $(q'_1,q'_2,q'_3)$. (b) The gnomonic projection of the point $(q'_1,q'_2,q'_3)$ from the sphere's center onto the point $(x,y)$ on the tangent plane.}
\label{Fig:1}
\end{figure}

We consider a spheroid enclosed by a sphere of radius $a$ and embedded in a three-dimensional Euclidean
space and choose a Cartesian coordinate system with the axes and the origin of the spheroid, $O_q$, as shown in Fig~\ref{Fig:1}.
If ${\vec r}$ denotes the vector of a point on the spheroidal surface designated by the Cartesian
coordinates $(q_{1}, q_{2},q_{3})$, these coordinates satisfy the implicit equation of the spheroid as follows:
 \begin{equation}\label{spheroid}
  \frac{q_{1}^{2}+q_{2}^{2}}{b^{2}}+\frac{q_{3}^{2}}{a^{2}}=1,
 \end{equation}
where $a$ and $b$ are, respectively, the polar radius and the equatorial radius of the spheroid, as seen in Fig~\ref{Fig:1} (a).

Let $ s$ be the arc-length of the geodesics from the north pole of the spheroid,
${\vec r}_{0}=(0,0,a)$, to the point ${\vec r}=(q_{1}, q_{2},q_{3})$. Therefore, we have
\be\label{the square of the distance}
  s^2={\cal G}_{\alpha\beta} q_{\alpha}q_{\beta}\ \ (\alpha,
  \beta=1,2),
\ee
where the metric tensor on spheroid is given by
\begin{eqnarray}\label{metric tensor on spheroid}
{\cal G}_{\alpha\beta}=\delta_{\alpha
\beta}+(\frac{a}{b})^{2}\frac{q_{\alpha}q_{\beta}}{b^{2}-q_{1}^{2}-q_{2}^{2}}.
\end{eqnarray}
Accordingly, the potential
energy for an isotropic harmonic oscillator of unit mass on the spheroidal surface
can be written as:
 \be\label{v}
  V_{\rm HO}(s)\equiv\frac{\omega}{2}s^{2}=\frac{\omega}{2}{\cal G}_{\alpha\beta} q_{\alpha}q_{\beta}
  =\frac{\omega}{2}(q_{1}^{2}+q_{2}^{2})\left[1+(\frac{a}{b})^{2}\frac{q_{1}^{2}+q_{2}^{2}}{b^{2}-q_{1}^{2}-q_{2}^{2}}
  \right].
 \ee
If we consider the following projection, which is projection from the spheroidal space to the spherical space,
\begin{eqnarray}\label{q prime}
q^{\prime}_{1}
&=&
\frac{a}{b}q_{1},\n\\
q^{\prime}_{2}
&=&
\frac{a}{b}q_{2},\n\\
q^{\prime}_{3}
&=&
q_{3},
\end{eqnarray}
it is obvious that the new Cartesian coordinates: $q^{\prime}_{1}, q^{\prime}_{2}$ and $q^{\prime}_{3}$, with the origin $O_q$ in the Fig.~\ref{Fig:1}, satisfy
the sphere equation as
\be
 {q^{\prime}_{1}}^2+{q^{\prime}_{2}}^2+{q^{\prime}_{3}}^2=1/\lambda^{2},
\ee
where $\lambda=1/a^2$ is the curvature of the enclosed sphere. Now, we use the two-dimensional gnomonic projection, which is the projection onto the
tangent plane from the center of the sphere in the embedding space. In this manner, the points on the sphere can be expressed in term of the coordinates of this projection.
Denoting the gnomonic projected Cartesian coordinates by $(x,y)$, with the origin $O_x$
which is also the point of tangency of the spheroid and the tangent plane, as shown in the Fig.~\ref{Fig:1} (b), the relation between the gnomonic projected coordinates and the point coordinates $(q'_1, q'_2,q'_3)$ on the sphere is given by
 \bea\label{q prime proj}
   q^{\prime}_{1}&=&\frac{a x}{\sqrt{a^{2}+\rho^{2}}},\n\\
   q^{\prime}_{2}&=&\frac{a y}{\sqrt{a^{2}+\rho^{2}}},\n\\
   q^{\prime}_{3}&=&\pm\frac{a^{2}}{\sqrt{a^{2}+\rho^{2}}}.
 \eea
Here, $\rho^{2}=x^{2}+y^{2}$ and the sign $+$ ($-$) in the last line refers to the point coordinates on
upper (lower) half of the sphere. Combining Eqs.~(\ref{q prime})
and (\ref{q prime proj}), one can obtain the projection of the coordinates $(q_1, q_2,q_3)$ of a point on the spheroid onto
the tangent plane as,
 \bea\label{q proj}
   q_{1}&=&\frac{b x}{\sqrt{a^{2}+\rho^{2}}},\n\\
   q_{2}&=&\frac{b y}{\sqrt{a^{2}+\rho^{2}}},\n\\
   q_{3}&=&\pm\frac{a^{2}}{\sqrt{a^{2}+\rho^{2}}}.
 \eea
In other word, the coordinates $(q_1, q_2,q_3)$ of a point on the
spheroid in term of the gnomonic projected coordinates are given by,
 \be
   {\vec
   r}=(\frac{bx}{\sqrt{a^{2}+\rho^{2}}},\frac{by}{\sqrt{a^{2}+\rho^{2}}},\pm\frac{a^{2}}{\sqrt{a^{2}+\rho^{2}}}).
 \ee
Let us take the total differential of ${\vec r}$ as,
 \be
   d {\vec r}={\vec r}_{x} dx+{\vec r}_{y} dy,
 \ee
where, ${\vec r}_{x}$ and ${\vec r}_{y}$ are, respectively, partial
derivatives of ${\vec r}$ with respect to the tangent plane coordinates $x$ and $y$.
Thus, after lengthy but straightforward
calculations, we obtain the metric of the spheroid as,
 \bea\label{ds}
   ds^{2}
   &\doteq&
   d{\vec r}\cdot d{\vec r}\n\\
   &=&
   \frac{1}{(1+\lambda \rho^{2})(1+\varepsilon)}\left\{d\vec{x}^{\,2}-\frac{\lambda(\vec{x}\cdot
   d\vec{x})^{2}}{1+\lambda \rho^{2}}
   \left(1-\frac{\varepsilon}{1+\lambda \rho^{2}}\right)   \right\},
 \eea
where ${\vec{x}}=(x,y)$ and $\varepsilon=\frac{a^{2}}{b^{2}}-1$ is the square of the second eccentricity of the spheroid that reflects the deviation from the sphericity.

In what follow, we consider a particle of unit mass constrained to move on the spheroidal surface.
From Eq.~(\ref{ds}), the kinetic energy of the particle can be written as
 \bea\label{sdot}
   T=\frac{1}{2}\left(\frac{ds}{dt}\right)^2
   =
   \frac{1}{2(1+\lambda \rho^{2})(1+\varepsilon)}\left\{\dot{\vec{x}}^{\,2}-\frac{\lambda(\vec{x}\cdot
   \dot{\vec{x}})^{2}}{1+\lambda \rho^{2}}
   \left(1-\frac{\varepsilon}{1+\lambda \rho^{2}}\right)   \right\}.
 \eea
With the help of Eq.~(\ref{q proj}), we can also represent the isotropic harmonic
oscillator potential on the spheroid, Eq. (\ref{v}), in terms of the tangent plane coordinates $x$ and $y$ as follows,
 \be\label{Vs}
  V_{\rm osc}(x,y)=\frac{\omega}{2}\ \frac{b^{2}+\rho^{2}}{a^{2}+\rho^{2}}\ \rho^{2}
  =\frac{\omega}{2}\ \frac{(1+\varepsilon)^{-1}+\lambda \rho^{2}}{1+\lambda \rho^{2}}\ \rho^{2}.
 \ee

Following the canonical quantization to describe the quantum dynamic behavior of the moving particle on the spheroidal background under the action of harmonic oscillator potential, we begin our analysis by considering the classical Lagrangian of the system in the tangent plane coordinates. To do so, we use Eqs.~(\ref{sdot}) and (\ref{Vs}) and yields
\bea\label{Lag}
   {\cal{L}^{\rm osc}}
   &\doteq&
   T-V_{\rm osc}(x,y)\n\\
   &=&
   \frac{1}{2}\ \frac{1}{(1+\lambda
   \rho^{2})(1+\varepsilon)}\left\{\dot{\vec{x}}^{2}-\frac{\lambda(\vec{x}\cdot
   \dot{\vec{x}})^{2}}{1+\lambda
   \rho^{2}}
   \left(1-\frac{\varepsilon}{1+\lambda \rho^{2}}\right)   \right\}\n\\
   &&-
   \frac{\omega}{2}\ \frac{(1+\varepsilon)^{-1}+\lambda \rho^{2}}{1+\lambda \rho^{2}}\ \rho^{2}\n\\
   &=&
   {\cal{L}}^{\rm osc}_{0}-\varepsilon\left\{{\cal{L}}_{0}-\frac{\lambda(\vec{x}\cdot
   \dot{\vec{x}})^{2}}{2(1+\lambda
   \rho^{2})^{3}}-\frac{\omega}{2}\left[\frac{1}{(1+\lambda\rho^{2})}+1\right]\rho^{2} \right\}\n\\
   &&+
   O(\varepsilon^{2}),
 \eea
where the Lagrangian
 \be
   {\cal{L}}^{\rm osc}_{0}=\frac{1}{2(1+\lambda
   \rho^{2})}\left[\dot{\vec{x}}^{2}-\frac{\lambda(\vec{x}\cdot
   \dot{\vec{x}})^{2}}{1+\lambda \rho^{2}}\right]-\frac{\omega}{2}\rho^{2}
   ,
 \ee
associated with the case that the particle in the presence of harmonic oscillator potential moving on a sphere with curvature $\lambda={1}/{a^{2}}$ , {\rm i.e.}, when the second eccentricity parameter is zero, $\varepsilon=0$. Here, $O(\varepsilon^{2})$ represents the terms of order equal
to or greater than  $\varepsilon^{2}$.

Using the Lagrangian~(\ref{Lag}), we can obtain the momentum conjugate for the variable $\vec{x}$ as
 \bea
   \vec{p}
   &=&
   \frac{\partial {\cal{L}}^{\rm osc}}{\partial \dot{x}}\hat{i}+\frac{\partial {\cal{L}}^{\rm osc}}{\partial
   \dot{y}}\hat{j}\n\\
   &=&
   \frac{1}{(1+\lambda \rho^{2})(1+\varepsilon)}\left[\dot{\vec{x}}-\frac{\lambda(\vec{x}\cdot
   \dot{\vec{x}})\vec{x}}{1+\lambda \rho^{2}}
   \left(1-\frac{\varepsilon}{1+\lambda \rho^{2}}\right)   \right]\n\\
   &=&
   {\vec{p}}_{0}-\varepsilon\left[{\vec{p}}_{0}-\frac{\lambda(\vec{x}\cdot
   \dot{\vec{x}})\vec{x}}{(1+\lambda \rho^{2})^{3}}\right]+O(\varepsilon^{2}),
 \eea
where the momentum vector of the system on the sphere is given by,
 \be\label{momentum vector of the system}
   {\vec{p}}_{0}=\frac{1}{(1+\lambda
   \rho^{2})}\left[\dot{\vec{x}}-\frac{\lambda(\vec{x}\cdot
   \dot{\vec{x}})\vec{x}}{1+\lambda \rho^{2}}
   \right].
 \ee

With proper calculations, the Hamiltonian of the two-dimensional isotropic
oscillator constrained to a spheroidal background is written as,
 \bea\label{hclass}
   H^{\rm osc}&\doteq&
   \dot{\vec{x}}\cdot{\vec{p}}-{\cal{L}}^{\rm osc}\\
   &=&
   H_{0}^{\rm osc}+H_{\varepsilon}^{\rm osc},
 \eea
where the effective potential $H^{\rm osc}_{\varepsilon}$ induced by the square of the second eccentricity of
the spheroid is given by
\begin{equation}\label{effective potential W}
H_{\varepsilon}^{\rm osc}=-\varepsilon\left\{H_{0}-\frac{\lambda}{2}\frac{({\vec{x}}\cdot{\dot{\vec{x}}})^{2}}{(1+\lambda\rho^{2})^{3}}
   +\frac{\omega}{2}\left[\frac{1}{(1+\lambda\rho^{2})}+1\right]\rho^{2}\right\}
   +
   O(\varepsilon^{2}),
\end{equation}
and  $ H_{0}^{\rm osc}$ is the well known Hamiltonian of a two-dimensional isotropic oscillator on a sphere with the curvature $\lambda={1}/{a^{2}}$ derived
by Higgs~\cite{Higgs1979},
 \be\label{Higgs Hamiltonian for an oscillator on a sphere}
   H_{0}^{\rm osc}=\frac{1}{2}(1+\lambda\rho^{2})\left[{p}^{2}_{0}
   +\lambda({\vec{x}\cdot{\vec{p}}_{0}})^{2} \right]
   +\frac{\omega}{2}\rho^{2}.
 \ee
Alternatively, the above Hamiltonian can be written in the form
\be\label{Higgs Hamiltonian for an oscillator on a sphere 2}
   H_{0}^{\rm osc}=\frac{1}{2}(\pi^{2}+\lambda L^{2})
   +\frac{\omega}{2}\rho^{2},
 \ee
where
\be\label{pi}
   \vec{\pi}={\vec{p}}_{0}+\lambda\vec{x}(\vec{x}.\vec{p}_{0}),
 \ee
and
\begin{eqnarray}\label{l}
\vec{L}=\vec{x}\times \vec{p}_{0}.
\end{eqnarray}
is the angular momentum of the system on the sphere.
For further calculation, we can introduce the following relation:
 \be
   {\vec{x}}\cdot{\dot{\vec{x}}}=({\vec{p}}_{0}\cdot{\vec{x}})(1+\lambda\rho^{2})^{2},
 \ee
which enable us to rewrite the effective potential~(\ref{effective potential W}) in the convenient form,
\bea\label{hclass new}
   H_{\varepsilon}^{\rm osc}
   &=&-\varepsilon\left\{H_{0}-\frac{\lambda}{2}({\vec{p}}_{0}\cdot{\vec{x}})^{2}(1+\lambda\rho^{2})
   +\omega\left[\frac{1}{(1+\lambda\rho^{2})}+1\right]\rho^{2}\right\}+
   O(\varepsilon^{2})\n\\
   &=&
   -\frac{\varepsilon}{2}\left\{(1+\lambda\rho^{2})\
   {p}_{0}^{2}+\omega\left[\frac{1}{(1+\lambda\rho^{2})}+1\right]\rho^{2}\right\}+O(\varepsilon^{2}).
 \eea
Let us consider the quantum mechanical description of the behavior of the isotropic oscillator on a spheroid. We first quantize the Hamiltonian (\ref{hclass})
by replacing classical position and momentum by related operators, and then impose equal-time commutation relation
among these variables, {\rm i.e}, $[\hat{x}_i,\hat{p}_j]=\imath\delta_{ij}\, (i,j=1,2)$. Thus, we have
\be\label{QH}
   \hat{H}^{\rm osc}=\hat{H}_{0}^{\rm osc}+\hat{H}_{\varepsilon}^{\rm osc}.
\ee
where
\begin{equation}\label{effective potential W}
\hat{H}_{\varepsilon}^{\rm osc}=-\frac{\varepsilon}{2}\left\{(1+\lambda\hat{\rho}^{2})\
   {\hat{p}}_{0}^{2}
   +\omega\left[1+\frac{1}{(1+\lambda\hat{\rho}^{2})}
   \right]\hat{\rho}^{2}\right\}+O(\varepsilon^{2}),
\end{equation}
and
\be\label{HO}
   \hat{H}_{0}^{\rm osc}=\frac{1}{2}(\hat{\pi}^{2}+\lambda \hat{L}^{2})
   +\frac{1}{2}\hat{\rho}^{2},
 \ee
are, respectively, the quantum counterpart of the classical Eqs.~(\ref{hclass new}) and~(\ref{Higgs Hamiltonian for an oscillator on a sphere 2}), in which the operators $\hat{\pi}$ and $\hat{L}$ are introduced by writing Eqs.~(\ref{pi})
and (\ref{l}) such that the operators $\vec{\hat{x}}$ and $\vec{\hat{p_0}}$ appear in symmetric order,
\begin{equation}\label{higgs 14a}
            \vec{\hat{\pi}}=\vec{\hat{p}}_{0}+\frac{\lambda}{2}\left[\vec{\hat{x}}(\vec{\hat{x}}\
            .\ \vec{\hat{p}}_{0})+(\vec{\hat{p}}_{0}\ .\
            \vec{\hat{x}})\vec{\hat{x}}\right],
\end{equation}
\begin{equation}\label{higgs 5.5}
            \hat{L}^{2}=\frac{1}{2}\hat{L}_{ij}\hat{L}_{ij},\h \hat{L}_{ij}
            =\hat{x}_{i}\hat{p}_{0j}-\hat{x}_{j}\hat{p}_{0 i}.
\end{equation}
For further calculations, it can be easily shown that the momentum operator of the system on the sphere, $\vec{\hat{p}}_{0}$, in the coordinate representation has the form,
\bea\label{realization}
   \vec{\hat{p}}_{0}&=&-\imath \frac{1}{\sqrt[4]{g(\rho)}}
   \vec{\nabla}\left(\sqrt[4]{g(\rho)}\ \ \cdot\ \right)=
   -\imath(\vec{\nabla}-\frac{3}{2}\ \frac{\lambda \vec{x}}{1+\lambda {\rho}^2}),
 \eea
which is identical to $-\imath\vec{\nabla}$ in the Euclidean space when the curvature $\lambda$ goes to zero.
It is worth noting that the Hermiticity condition for a generic operator $\mathcal{O}$ on a sphere is given by~\cite{Leemon1979}
\be\label{Hermiticity condition}
   \int_{\rm sphere}\sqrt{g(\rho)}\ d\vec{x}\ \psi^{\ast}(\vec{x})\mathcal{O}\varphi(\vec{x})
   =\int_{\rm sphere}\sqrt{g(\rho)}\ d\vec{x}\
   \varphi(\vec{x})\left(\mathcal{O}\psi(\vec{x})\right)^{\ast},
 \ee
where $\psi(\vec{x})$ and $\varphi(\vec{x})$ are the wave functions describing motion of the particle on the sphere, and $g(\rho)=(1+\lambda \rho^{2})^{-3},$
denotes the determinant of the metric of the sphere.
For further convenience, we chose these wave functions such that to satisfy the invariant normalization condition,
\be\label{Norm}
   \int_{\rm sphere}\sqrt{g(\rho)}\ d\vec{x}\ \psi^{\ast}(\vec{x})\psi(\vec{x})
   =1.
\ee

%%%%%%%%%%%%%%%%%%%%%%%%%%%%%%%%%%%%%%%%%%%%%%%%%%%%%%%%%%%%%%%%%%%%%%%%%%%%%%%%%%%%%%%%%%%%%%%%%%%%%%%%%%%%%%%%%%%%%%%%%%%%%
\section{Free particle on a spheroid}\label{FPS}
Let us start our analysis by considering the eigenvalues and eigenfunctions of a free particle of unit mass constrained to move on the spheroidal surface. To this end, we use Eq.~(\ref{realization}) and express the Hamiltonian~(\ref{QH}) in the coordinate representation and then set $\omega=0$. Thus, it yields
\bea\label{Hamiltonian of the free particle on a sphroid}
   \hat{H}^{{\rm free}}
   &=&\hat{H}_{0}^{{\rm free}}+\hat{H}_{\varepsilon}^{{\rm free}},
\eea
where
\bea\label{Hamiltonian of the free particle on a sphere}
   \hat{H}_{0}^{{\rm free}}
   &=&-\frac{1}{2}(1+\lambda {\rho}^{2})\left[\vec{\nabla}^{2}+\lambda(\vec{x}.\vec{\nabla})^{2}
   +\lambda(\vec{x}.\vec{\nabla})\right]\\
   &=&
   -\frac{1}{2}(1+\lambda {\rho}^{2})\left[\frac{1}{\rho}(1+2\lambda\rho^{2})\frac{\partial}{\partial
   \rho}
   +(1+\lambda {\rho}^{2})\frac{\partial^{2}}{\partial \rho^{2}}
   +\frac{1}{\rho^{2}}\frac{\partial^{2}}{\partial \varphi^{2}}
   \right],\n
\eea
is the Hamiltonian of the free particle on a sphere in the polar coordinates $(\rho,\varphi)$ of the tangent plane,
%, {\rm i.e.}, by using the relation $(x,y) = (\rho \cos \varphi ,\rho \sin \varphi)$
while the effective potential $\hat{H}_{\varepsilon}^{{\rm free}}$  in this coordinate has the form
\begin{equation}\label{effective potential for free paticle}
\hat{H}_{\varepsilon}^{{\rm free}}=-\frac{\varepsilon}{2}
   (1+\lambda {\rho}^{2}) \hat{p}_{0}^{2}+O(\varepsilon^{2}).
\end{equation}
We see from Eq.~(\ref{Hamiltonian of the free particle on a sphroid}) that the problem of the motion of the free particle on the spheroid reduces to the finding of the eigenvalues and eigenfunctions of the Hamiltonian $\hat{H}_{0}^{{\rm free}}$ with the effective potential $\hat{H}_{\varepsilon}^{{\rm free}}$. The quantum
dynamics the free motion of the particle on the sphere is basically governed by the Schr\"{o}dinger equation,
\be
 \hat{H}_{0}^{{\rm free}}\psi_{n,{\rm free}}^{(0) }(\rho,\varphi)
 =E_{n,{\rm free}}^{(0) }\psi_{n,{\rm free}}^{(0) }(\rho,\varphi).
\ee
By separating the wave functions into a radial
part and an angular part, the eigenfunctions and eigenvalues can be obtained finally as
\bea
  \psi_{n,{\rm free}}^{(0) }(\rho,\varphi)&=& a_{n}(\frac{\lambda {\rho}^{2}}{1+\lambda
  {\rho}^{2}})^{\frac{n}{2}}e^{\imath n \varphi},\n\\
  E_{n,{\rm free}}^{(0) }&=& \frac{\lambda}{2}n(n+1),
\eea
where $a_{n}$ is a normalization coefficient. To normalize the eigenfunction $\psi_{n,{\rm free}}^{(0) }(\rho,\varphi)$, we use the polar coordinates~(\ref{Norm}) and arrive at
\bea\label{Norm1}
   a_{n}^{2}\int_{0}^{2 \pi}d\varphi\int_{0}^{\infty}d\rho\sqrt{g(\rho)}\ (\frac{\lambda
   {\rho}^{2}}{1+\lambda {\rho}^{2}})^{n}
   =a_{n}^{2}\frac{2 \pi}{\lambda}\int_{0}^{\frac{\pi}{2}}d\chi(\sin\chi)^{2n}
   =1.
\eea
Here, in the second equality, the change of variable $\sin\chi=(\frac{\lambda
{\rho}^{2}}{1+\lambda {\rho}^{2}})^{\frac{1}{2}}$ is used.
Now, by using the following integral identity
\bea\label{}
   \int_{0}^{\frac{\pi}{2}}d\chi(\sin\chi)^{n}
   =\frac{\pi^{\frac{1}{2}} \Gamma[\frac{1+n}{2}]}{2 \Gamma[1+\frac{n}{2}] },
\eea
we obtain the normalization constant as
\bea\label{normalization constant}
   a_{n}=\sqrt{\frac{\lambda\ \Gamma[\frac{3}{2}+n]}{\pi^{\frac{3}{2}}\ \Gamma[1+n]}},
\eea
where $\Gamma$ denotes the Gamma function.
Let us assume that the spheroid is almost a sphere of radius $a$.
In this case the the square of the second eccentricity $\varepsilon$ is small, and we can drop all the higher-order terms in~(\ref{effective potential for free paticle}) that contain the factor $\varepsilon^2$ and the power higher than 2.  As a consequence, the effective potential $\hat{H}_{\varepsilon}^{{\rm free}}$ can be considered as the small perturbation. In this manner, by using the nondegenerate perturbation theory, the first-order energy shift can be calculated as follows:
\be
    \Delta E_{n,{\rm free}}^{(1)}=-\frac{\varepsilon}{2}
    \langle \psi_{n,{\rm free}}^{(0) }| (1+\lambda {\rho}^{2}) \hat{p}_{0}^{2}
    |\psi_{n,{\rm free}}^{(0) }  \rangle.
\ee
In the coordinate representation, by making use of the polar coordinates $(\rho,\varphi)$ and the symmetrization of the above relation, we get
\bea
    \Delta E_{n,{\rm free}}^{(1)}
    &=&-\frac{\varepsilon}{4}
    \int_{0}^{2 \pi}d\varphi\int_{0}^{\infty}d\rho\sqrt{g(\rho)}\n\\
    &&\times
    \Big\{\psi_{n,{\rm free}}^{(0) \ast}(\rho,\varphi)\,(1+\lambda {\rho}^{2})
    \hat{p}_{0}^{2}\,\psi_{n,{\rm free}}^{(0) }(\rho,\varphi)\n\\
    &&+
    \psi_{n,{\rm free}}^{(0) \ast}(\rho,\varphi)\,\hat{p}_{0}^{2}(1+\lambda {\rho}^{2})\,
    \psi_{n,{\rm free}}^{(0) }(\rho,\varphi)\Big\}\n\\
    &=&
    -\frac{\varepsilon}{4}
    \int_{0}^{2 \pi}d\varphi\int_{0}^{\infty}d\rho\sqrt{g(\rho)}\n\\
    &&\times
    \bigg\{\left[\vec{\hat{p}}_{0}(1+\lambda
    {\rho}^{2})\psi_{n,{\rm free}}^{(0) }(\rho,\varphi)\right]^{\ast}
    .\left[\vec{\hat{p}}_{0}\psi_{n,{\rm free}}^{(0) }(\rho,\varphi)\right]\n\\
    &&+
    \left[\vec{\hat{p}}_{0}\psi_{n,{\rm free}}^{(0) }(\rho,\varphi)\right]^{\ast}
    .\left[\vec{\hat{p}}_{0}(1+\lambda
    {\rho}^{2})\psi_{n,{\rm free}}^{(0) }(\rho,\varphi)\right]\bigg\}.\n
\eea
By make use of Eq. (\ref{realization}) and performing the change of
variable $\sin\chi=(\frac{\lambda {\rho}^{2}}{1+\lambda {\rho}^{2}})^{\frac{1}{2}}$, after some algebra we find
\bea
    \Delta E_{n,{\rm free}}^{(1)}
    &=&-\pi \varepsilon a_{n}^{2}
    \int_{0}^{\frac{\pi}{2}}d\chi \, \sin\chi \Big\{2 n^{2}(\sin\chi)^{2n-2}\\
    &&+
    (n^{2}+n+\frac{9}{4}-3)(\sin\chi)^{2n+2}-
    n(2n+1)(\sin\chi)^{2n}
     \Big\}.\n
\eea
Finally, by using the normalization constant~(\ref{normalization constant}) and performing the integration over $\chi$, the first-order energy shift is obtained as
\bea\label{first-order energy shift}
    \Delta E_{n,{\rm free}}^{(1)}&=&\frac{\varepsilon \lambda}{2}\Bigg\{n(2n+1)-2 n^{2}\frac{\Gamma[n]
    \Gamma[\frac{3}{2}+n]}{\Gamma[1+n] \Gamma[\frac{1}{2}+n]  } \nonumber\\
    &&-(n^{2}+n-\frac{3}{4})\frac{\Gamma[2+n]  \Gamma[\frac{3}{2}+n]}{\Gamma[1+n] \Gamma[\frac{5}{2}+n]  }
    \Bigg\}.
\eea
The analysis of the first order energy shift~(\ref{first-order energy shift}) shows that it depends on the quantum number $n$, the curvature $\lambda$ and the square of the second eccentricity $\varepsilon$.

For the eigenstates describing the free motion of the particle on the spheroid, one can write in the first order of the perturbation theory that
\bea
 |\psi_{n,{\rm free}}^{(1) }\rangle=-\frac{\varepsilon}{2}\sum_{m\neq n}\frac{\langle
 \psi_{m,{\rm free}}^{(0) }|
(1+\lambda {\rho}^{2}) \hat{p}_{0}^{2}| \psi_{n,{\rm free}}^{(0)}
\rangle}{E_{n}-E_{m}} |\psi_{n,{\rm free}}^{(0)}\rangle.
\eea
In the polar coordinate representation, by making use of the symmetrization of the above relation we arrive at
\bea
  \psi_{n,{\rm free}}^{(1) }(\rho,\varphi) &=&-\frac{\varepsilon}{4}\sum_{m\neq
  n}\psi_{m, {\rm free}}^{(0)}(\rho,\varphi)
  \int_{0}^{2 \pi}d\varphi\int_{0}^{\infty}d\rho\sqrt{g(\rho)}\n\\
  &&\times
  \Big\{\psi_{m, {\rm free}}^{(0) \ast}\,(\rho,\varphi)(1+\lambda {\rho}^{2})
  \hat{p}_{0}^{2}\, \psi_{n, {\rm free}}^{(0)}(\rho,\varphi)\n\\
  &&-
  \psi_{m, {\rm free}}^{(0) \ast}\,(\rho,\varphi)\hat{p}_{0}^{2}(1+\lambda {\rho}^{2})\,
    \psi_{n,{\rm free}}^{(0)}(\rho,\varphi)\Big\}\n\\
  &=&
  0.
\eea
%
%Consequently, we find that the second order energy shift becomes zero.
Furthermore, it can be easily shown that the second-order energy shift is also zero.

%%%%%%%%%%%%%%%%%%%%%%%%%%%%%%%%%%%%%%%%%%%%%%%%%%%%%%%%%%%%%%%%%%%%%%%%%%%%%%%%%%%
%%%%%%%%%%%%%%%%%%%%%%%%%%%%%%%%%%%%%%%%%%%%%%%%%%%%%%%%%%%%%%%%%%%%%%%%%%%%%%%%%%%
\section{An isotropic harmonic oscillator on a spheroid}\label{IHO}

In this section, we are interested to obtain an approximately expression for  the eigenvalues and the eigenfunctions of a particle moving in a spheroidical surface under the action of an isotropic harmonic oscillator
potential. Let us assume that $\varepsilon$ is small enough so that we can apply first-order perturbation theory. In this case, from Eq.~(\ref{QH}), we can imagine a perturbation as:
\begin{equation}
\hat{H}_{\varepsilon}^{\rm osc}=-\frac{\varepsilon}{2}\left\{(1+\lambda\hat{\rho}^{2})\
   {\hat{p}}_{0}^{2}
   +\omega\left[1+\frac{1}{(1+\lambda\hat{\rho}^{2})}
   \right]\hat{\rho}^{2}\right\},
\end{equation}
while the unperturbed Hamiltonian~(\ref{HO}) is the Hamiltonian of an isotropic oscillator on a sphere whose eigenvalues and eigenfunctions are known and given in~\cite{Leemon1979} as:
\bea\label{eigenvalues}
  E_{n,{\rm osc}}^{(0) }=(n+1)\Omega+\frac{\lambda}{2}(n+1)^{2},
\eea
and
\bea\label{eigenfunc}
  \psi_{nl,{\rm osc}}^{(0)}(\chi,\varphi)=\varphi_{nl,{\rm osc}}^{(0)}(\chi)\ e^{\imath l\varphi},
\eea
where
\bea
   \varphi_{nl,{\rm osc}}^{(0)}(\chi)
   =
   \mathcal{N}(\sin \chi)^{l}(\cos \chi)^{\beta+\frac{1}{2}}
   {_{2}F}_{1}\left(-k
   ,-\frac{1}{2}(n-l);l+1;\sin^{2} \chi\right).
\eea
Here, $l=-n,-n+2,\cdots,n$ and the integer $n$ is nonnegative, ${_{2}F}_{1}$ is the ordinary hypergeometric function in which
$
  k=-\beta-\frac{1}{2}(n+l)-1
$
and
$
  \Omega=\lambda\beta=\sqrt{\omega^{2}+\frac{\lambda^{2}}{4}}.
$
Furthermore, the normalization coefficient $\mathcal{N}$ is given by
\bea
  \mathcal{N}
  =
  \left[\frac{2\lambda \Gamma(k+l+1)
  \ \Gamma(k+l+\beta+1)\ (2k+l+\beta+1)}{k!\ \Gamma^{2}(l+1)\ \Gamma(k+\beta+1)}\right]^{\frac{1}{2}}.
\eea
As it is seen from Eqs. (\ref{eigenvalues}) and (\ref{eigenfunc}), the eigenfunctions $\psi_{nl,{\rm osc}}^{(0)}(\chi,\varphi)$ are degenerate. But, it is worth noting that in the degenerate perturbation theory, if the perturbation is also diagonal in the used representation, all we need to do for the first-order energy shift is to calculate the expectation value as follows \cite{Sakurai2010},
\bea\label{first-order energy shift for oscillator}
    \Delta E_{n,{\rm osc}}^{(1)}
    =
    -\frac{\varepsilon}{2}
    \langle \psi_{nl,{\rm osc}}^{(0)}| \left\{(1+\lambda\hat{\rho}^{2})\ {\hat{p}}_{0}^{2}
    +\omega\left[\frac{1}{(1+\lambda\hat{\rho}^{2})}
    +1\right]\hat{\rho}^{2}\right\} |\psi_{nl,{\rm osc}}^{(0)}  \rangle.
\eea
Recalling the Hermiticity condition~(\ref{Hermiticity condition}), the first term of the above relation can be calculated as
\bea\label{55}
    &-&
    \frac{\varepsilon}{2}
    \langle \psi_{nl,{\rm osc}}^{(0)}| (1+\lambda\hat{\rho}^{2})\ {\hat{p}}_{0}^{2}
    |\psi_{nl,{\rm osc}}^{(0)} \rangle\n\\
    &=&
    -\frac{\varepsilon}{4}
    \int_{0}^{2 \pi}d\varphi\int_{0}^{\infty}d\rho\sqrt{g(\rho)}\n\\
    &&\times
    \{[-\imath(\vec{\nabla}-\frac{3}{2}\ \frac{\lambda \vec{x}}{1+\lambda {\rho}^2})(1+\lambda
    {\rho}^{2})\psi_{nl,{\rm osc}}^{(0)}(\rho,\varphi)]^{\ast}\n\\
    &&.
    [-\imath(\vec{\nabla}-\frac{3}{2}\ \frac{\lambda \vec{x}}{1+\lambda
    {\rho}^2})\psi_{nl,{\rm osc}}^{(0)}(\rho,\varphi)]\n\\
    &&+
    [-\imath(\vec{\nabla}-\frac{3}{2}\ \frac{\lambda \vec{x}}{1+\lambda
    {\rho}^2})\psi_{nl,{\rm osc}}^{(0)}(\rho,\varphi)]^{\ast}\n\\
    &&.
    [-\imath(\vec{\nabla}-\frac{3}{2}\ \frac{\lambda \vec{x}}{1+\lambda {\rho}^2})(1+\lambda
    {\rho}^{2})\psi_{nl,{\rm osc}}^{(0)}(\rho,\varphi)]\}\n\\
    &=&
    -\frac{\varepsilon\pi}{\lambda}\int_{0}^{\frac{\pi}{2}}\sin\chi \,d\chi
        \bigg\{\lambda\cos^{2}\chi \Big[\frac{\partial\varphi_{nl,{\rm osc}}^{(0)}(\chi)}{\partial\chi}\Big]^{2}\n\\
    &&-
    \lambda\cos\chi\,\sin\chi\,\varphi_{nl,osc}^{(0)}(\chi)
    \frac{\partial\varphi_{nl,{\rm osc}}^{(0)}(\chi)}{\partial\chi}\n\\
    &&-
    \Big(\frac{3}{4}\lambda\sin^{2}\chi+\frac{\lambda
    l^{2}}{\sin^{2}\chi}\Big)[\varphi_{nl,{\rm osc}}^{(0)}(\chi)]^{2}\bigg\}.
\eea
By using the change of variable $x=1-2\sin^{2}\chi$, we can write:
\bea
   \frac{\partial\varphi_{nl,{\rm osc}}^{(0)}(\chi)}{\partial\chi}
   &=&
   \frac{\partial x}{\partial\chi}
   \frac{\partial\varphi_{nl,{\rm osc}}^{(0)}(x)}{\partial x}
   =
   -2(1-x)^{\frac{1}{2}}
   (1+x)^{\frac{1}{2}}\n\\
   &&\times
   \frac{\partial}{\partial x}\left[\mathcal{N}(\frac{1-x}{2})^{\frac{l}{2}}
   (\frac{1+x}{2})^{\frac{\beta}{2}+\frac{1}{4}}\frac{k!\Gamma(l+1)}{\Gamma(k+l+1)}
   P_{k}^{l,\beta}(x)\right]\n\\
   &=&
   -\frac{\mathcal{N}}{2^{(\frac{l}{2}+\frac{\beta}{2}-\frac{3}{4})}}
   \frac{k!\,\Gamma(l+1)}{\Gamma(l+1+k)}
   \bigg[-\frac{l}{2}
   (1-x)^{\frac{l-1}{2}}
   (1+x)^{\frac{\beta}{2}+\frac{3}{4}} P_{k}^{l,\beta}(x)\n\\
   &&+
   \frac{1}{4}(2\beta+1)
   (1-x)^{\frac{l+1}{2}}
   (1+x)^{\frac{\beta}{2}-\frac{1}{4}}P_{k}^{l,\beta}(x)\n\\
   &&+
   (1-x)^{\frac{l+1}{2}}
   (1+x)^{\frac{\beta}{2}+\frac{3}{4}}\frac{\Gamma(l+\beta+k+2)}{2\Gamma(l+\beta+k+1)}
   P_{k-1}^{l+1,\beta+1}(x)\bigg]\n\\
   &\equiv&
   \Phi_{n,l}(x),
\eea
where we have used the following relation between the hypergeometric function and the Jacobi polynomials \cite{Abramowitz}
\bea
  {_{2}F}_{1}\left(-k,l+\beta+k+1;l+1;\frac{1-x}{2}\right)
  =\frac{k!\,\Gamma(l+1)}{\Gamma(k+l+1)} P_{k}^{l,\beta}(x),
\eea
and also the identity \cite{Bell}
\bea
  \frac{d^{m}}{dx^{m}}\big(P_{k}^{\alpha,\beta}(x)\big)=
  \frac{\Gamma(\alpha+\beta+k+1+m)}{2^{m}\Gamma(\alpha+\beta+k+1)}
  P_{k-m}^{\alpha+m,\beta+m}(x).
\eea
Now, with the help of above equations, Eq.~(\ref{55}) can be recast as
\bea\label{first term}
  &-&
    \frac{\varepsilon}{2}
    \langle \psi_{nl,{\rm osc}}^{(0)}| (1+\lambda\hat{\rho}^{2})\ {\hat{p}}_{0}^{2}
    |\psi_{nl,{\rm osc}}^{(0)} \rangle\n\\
  &=&
  \frac{\varepsilon\pi}{2}
  \int_{-1}^{1}\frac{\frac{dx}{2}}{(\frac{1+x}{2})^{\frac{1}{2}}}\n\\
  &  &\times
 \bigg \{4\Big(\frac{1+x}{2}\Big)^{2}\Big(\frac{1-x}{2}\Big)[\Phi_{n,l}(x)]^{2}\n\\
  &  &-
  \Big(\frac{1+x}{2}\Big)^{\frac{1}{2}}\Big(\frac{1-x}{2}\Big)^{\frac{1}{2}}
  [\varphi_{n,l}(x)\Phi_{n,l}(x)]\n\\
  &  &-
  \Big[\frac{3}{4}\big(\frac{1-x}{2}\big)-\frac{2 l^{2}}{1-x}\Big][\varphi_{n,l}(x)]^{2}\bigg\}.
\eea
Similarly, the second term of the Eq. (\ref{first-order energy shift for oscillator}) can be calculated as
\bea\label{second term}
    &-&
    \frac{\varepsilon \omega}{2}
    \langle \psi_{nl,{\rm osc}}^{(0)}| [\frac{1}{(1+\lambda\hat{\rho}^{2})}
    +1]\hat{\rho}^{2}
    |\psi_{nl,{\rm osc}}^{(0)} \rangle\n\\
    &=&
    -\frac{\varepsilon\omega\pi}{\lambda^{2}} \int_{0}^{\frac{\pi}{2}} \sin\chi \, d\chi
    \big(\sin^{2}\chi
    +\tan^{2}\chi \big)\varphi_{n,l}^{2}(\chi) \n\\
    &=&
    -\frac{\varepsilon\omega\pi}{\lambda^{2}} \mathcal{N}^{2}
    \int_{0}^{\frac{\pi}{2}} \sin\chi \, d\chi
    \big(\sin^{2}\chi
    +\tan^{2}\chi\big)\n\\
    &&\times
    (\sin \chi)^{2l}(\cos \chi)^{2\beta+1}
    {_{2}F}_{1}^{2}\left(-k,-\frac{1}{2}(n-l);l+1;\sin^{2} \chi\right).
\eea
Due to the complexity of the final form of Eqs.~(\ref{first term}) and~(\ref{second term}), we
do not attempt to obtain the analytic form of Eq.~(\ref{first-order energy shift for oscillator}). Instead,
we numerically study the first-order energy shift for the oscillator on the spheroid.

 \begin{figure*}[ht]
 \begin{minipage}[b]{0.32\linewidth}
 \centering
 \includegraphics[width=\textwidth]{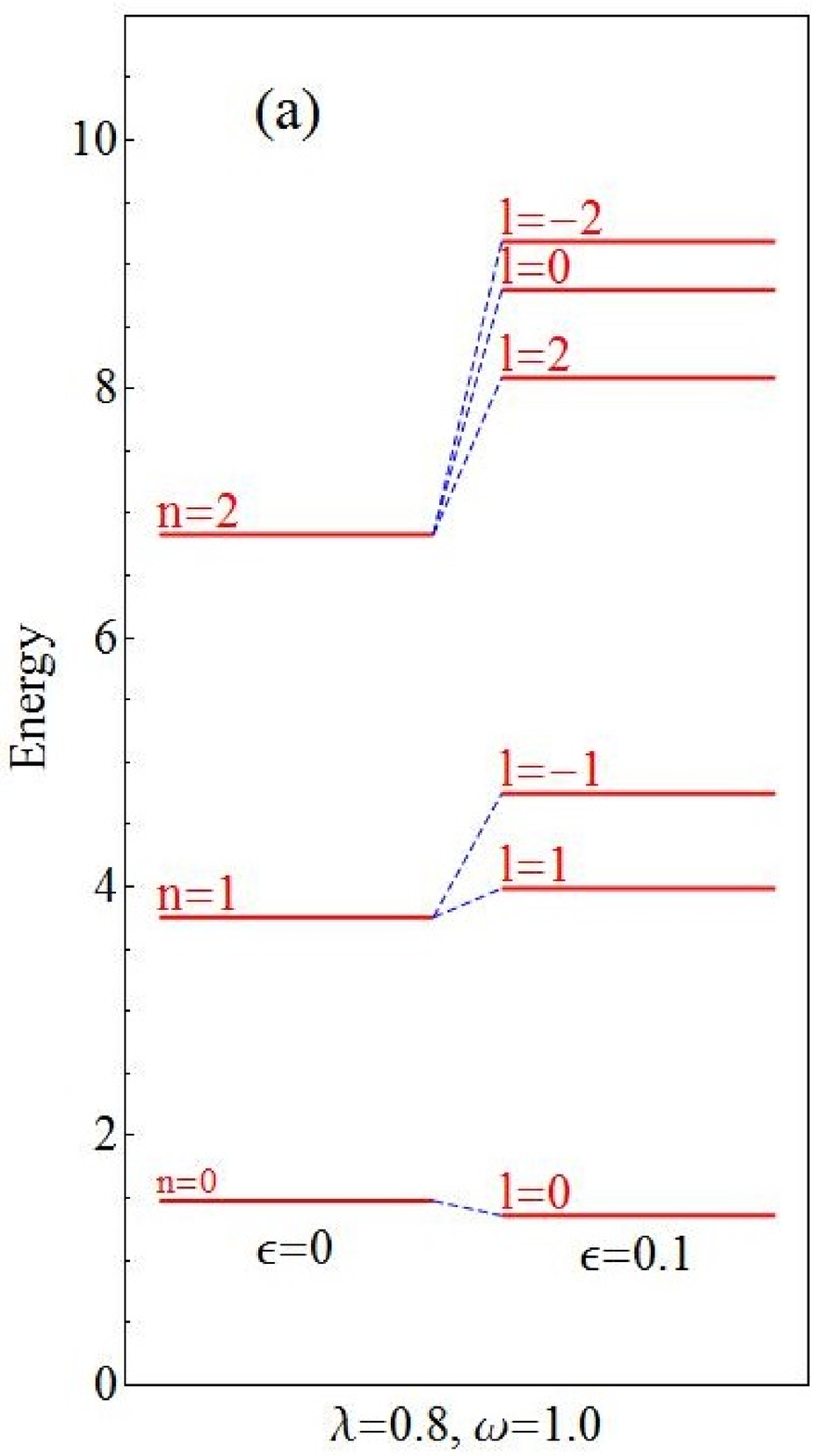}
 \end{minipage}
 \hspace{0cm}
 \begin{minipage}[b]{0.32\linewidth}
 \centering
 \includegraphics[width=\textwidth]{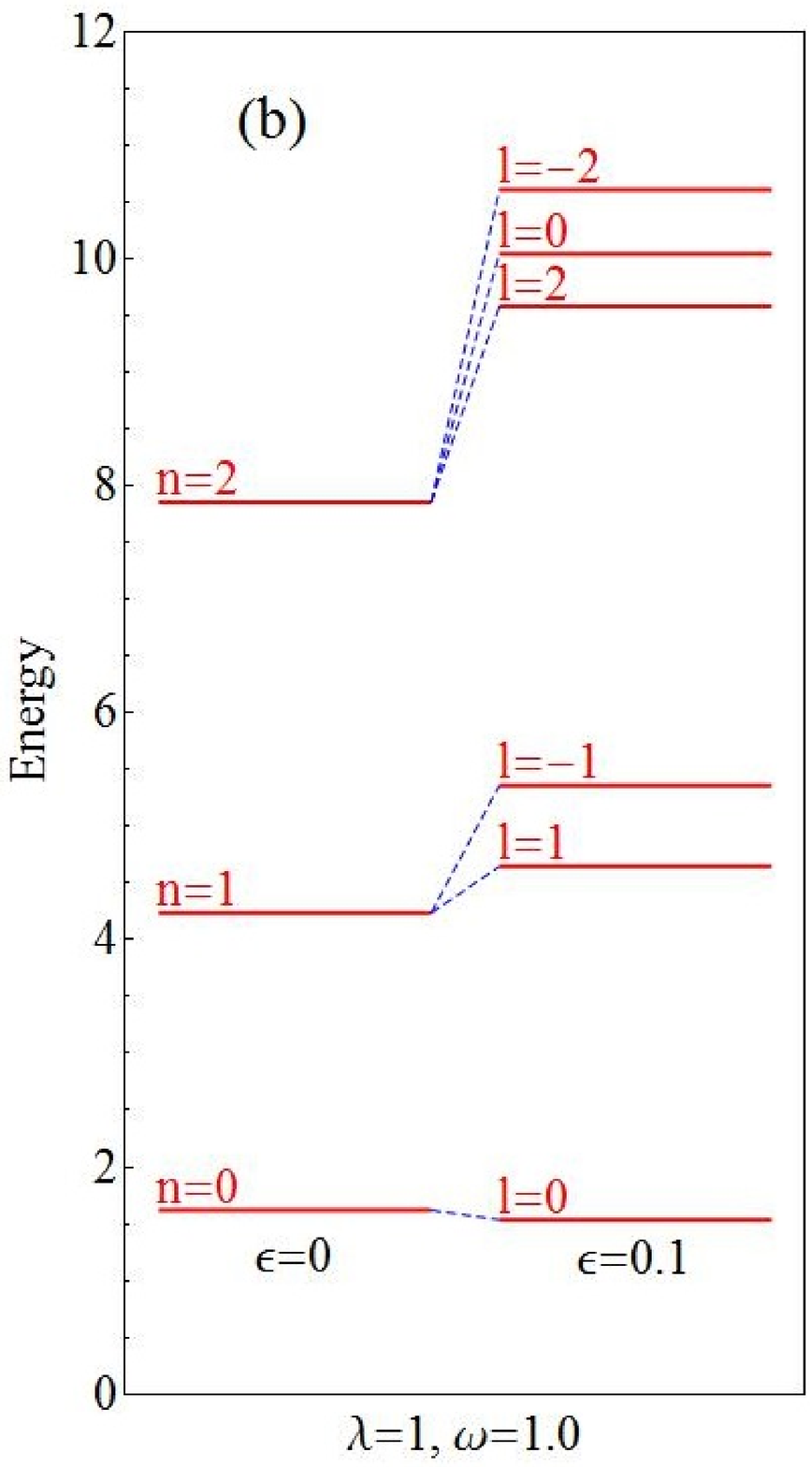}
 \end{minipage}
 \hspace{0cm}
 \begin{minipage}[b]{0.32\linewidth}
 \centering
 \includegraphics[width=\textwidth]{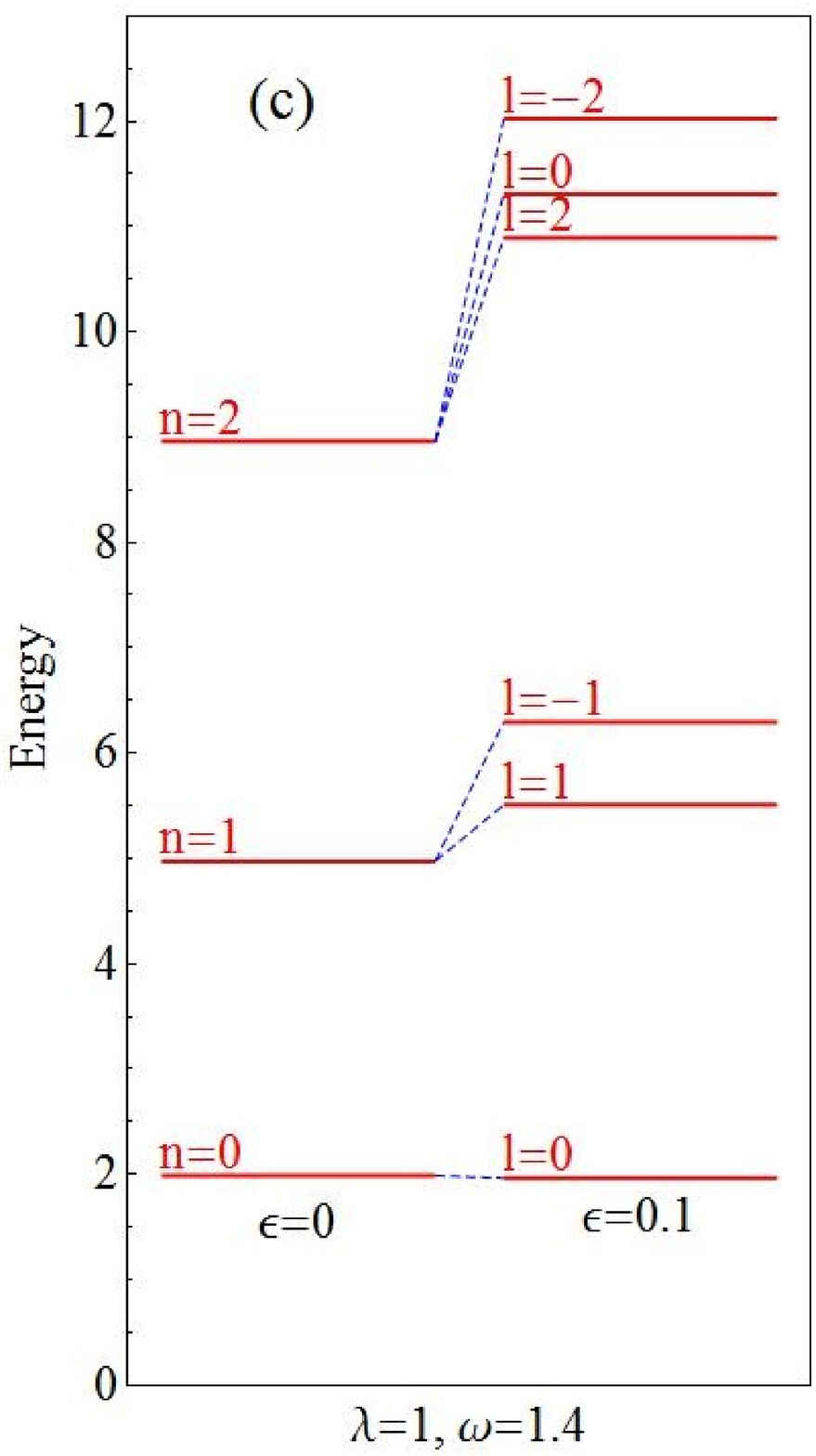}
 \end{minipage}
  \caption{Schematic energy-level diagram for the isotropic oscillator on a sphere ($\varepsilon=0$) and splitting its energy levels due to  the presence of small amounts of $\varepsilon=0.1$. Here, the parameters are chosen as: (a) $ \lambda=0.8$ and $\omega=1$, (b) $ \lambda=1.0$ and $\omega=1.0$, and (c) $\lambda=1.0$ and $\omega=1.4$.}\label{2}
 \end{figure*}

In Fig. \ref{2}, we have schematically plotted the energy levels of the isotropic oscillator on both the sphere ($\varepsilon=0$) and the spheroid. It is seen that the effect of nonzero but small value of $\varepsilon$ on the electronic states is to split the energy levels of degenerate states into $(n + 1)$ sublevels, which corresponds to the $(n +1)$ values of $l: l= -n, -n + 2, ...,n$, i.e., lifting the degeneracy.
%Unlike the energy levels of the isotropic oscillator on a sphere, Eq. (\ref{eigenvalues}),
%the energy levels of this system on a spheroid are not degenerate and each energy level is split
%
We also see that the energy splittings are not constant for different $l$. This is a direct consequence of Eqs.~(\ref{first term}) and~(\ref{second term}) which state that the spacing between the sublevels depend on the parameters $\varepsilon$, $\lambda$ and $\omega$. Comparing Figs.~\ref{2}(a) with~\ref{2}(b), we
observe that the splitting distance between the sublevels are reduced by increasing $\lambda$. On contrary, as the frequency increases, the spacing between the sublevels decreases (Fig.~\ref{2}(c)).

\section{Summary and Concluding Remarks}\label{summary}
In this paper, by using two consecutive projections, first from the spheroidal to the spherical space and then from the spherical space onto the tangent plane via the gnomonic projection, we have obtained the Hamiltonian of an isotropic harmonic oscillator confined to a spheroidal background in terms of the Cartesian coordinates of the tangent plane.
The aforementioned Hamiltonian has been quantized by replacing the position and momentum by the standard quantum mechanical position and momentum operators, and
then imposing equal-time commutation relation among these variables.
We have calculated the eigenfunctions and eigenvalues of a free particle on a sphere. Within the perturbation theory, we have approximately determined the eigenvalues and the eigenfunctions of a free particle on the spheroidal surface. The results show that the first order energy shift of the free particle on the spheroid depends on the quantum number $n$, the curvature $\lambda$ and the square of the second eccentricity $\varepsilon$.
Subsequently, we have determined the eigenvalues and the eigenfunctions for an isotropic harmonic oscillator problem on a spheroidal surface by using
the perturbation theory up to the first order in second eccentricity of the spheroid.
The obtained numerical results illustrate that the deviation from the sphericity ($\varepsilon\neq 0$) removes the degeneracy of the energy states associated with the problem of an isotropic oscillator on a sphere, and induces the splitting of energy levels into $(n + 1)$ sublevels related to the $(n +1)$ values of $l$. Moreover, we found that the splitting distance between the sublevels are increased(decreased) by increasing the angular frequency of the oscillator $\omega$(the curvature $\lambda$).

\section{Acknowledgment}
The authors wish to thank the Shahrekord University for their support.
A.M. also wishes to thank The Office of Graduate Studies and Research Vice President of The
University of Isfahan for their support.
%==================================================================================================
 
 \vspace{1cm}

%==================================================================================================

\end{document}